\documentclass[preprint,aps,showpacs,prb,floatfix]{revtex4}

\usepackage{amsmath}
\usepackage{graphicx}
\usepackage{subfigure}		
\usepackage{dcolumn}
\usepackage{bm}
\usepackage{color}
\usepackage[utf8]{inputenc}	
\usepackage{ulem}	

\makeatletter

\begin{document}
\title{Designing  lattice structures with maximal nearest-neighbor entanglement}
\author{J. C. Navarro-Mu\~noz$^1$, R. L\'opez-Sandoval$^1$, and   
M. E. Garc\'{\i}a$^2$ }
\affiliation{$^1$Instituto Potosino de Investigaci\'on Cient\'{\i}fica
y Tecnol\'ogica, Camino a la presa San Jos\'e 2055, 78216
San Luis Potos\'{\i}, Mexico}
\affiliation{$^2$   Theoretische Physik, FB 18, Universit\"at Kassel and 
 Center for Interdisciplinary Nanostructure Science and Technology 
(CINSaT),
Heinrich-Plett-Str.40, 34132 Kassel, Germany}

\begin{abstract} 
In this work, we study the numerical optimization of
nearest-neighbor concurrence of bipartite one and two dimensional
lattices, as well as non bipartite two dimensional lattices. These
systems are described in the framework of a tight-binding Hamiltonian
while the optimization of concurrence was performed using genetic
algorithms. Our results show that the concurrence of the optimized
lattice structures is considerably higher than that of non optimized systems. In
the case of one dimensional chains the concurrence is
maximized when the system begins to dimerize, i.e. it undergoes a
structural phase transition (Peierls distortion).  This result is
consistent with the idea that entanglement is maximal or shows a
singularity near quantum phase transitions and that quantum
entanglement cannot be freely shared between many objects (monogamy
property).  Moreover, the optimization of concurrence in
two-dimensional bipartite and non bipartite lattices is achieved when
the structures break into smaller subsystems, which are arranged in
geometrically distinguishable configurations.  This behavior is again
related to the monogamy property.

\end{abstract}

\maketitle

\section{Introduction} 
Quantum entanglement is one of the most
distinctive features in quantum mechanics and yet its properties are still not
fully understood. This quantum resource is considered a key element
for several quantum information and quantum computation proposals such
as quantum teleportation \cite{teleportation}, superdense coding
\cite{superdensecoding}, certain kinds of quantum key distribution
schemes and quantum secret sharing protocols \cite{Brassard,Nielsen}.

Recently, much research has been focused on a better understanding of
quantum correlations in multiparticle systems
\cite{Connor,Wootters,Dur,Koashi}. A characteristic property that sets
apart quantum correlations (or entanglement) from the classical ones is
that entanglement cannot be freely shared among many objects. For the
special case of three qubits $A$, $B$ and $C$, this has been shown by
Coffman et al. \cite{Coffman} using concurrence, a measure of
entanglement taking values between $0$ and $1$ \cite{Wootters1}. Coffman
et al. showed that the sum of the squared concurrences $AB$ and $BC$
cannot be greater than unity, meaning that the degree of entanglement
between $A$ and $B$ limits the entanglement between $A$ and $C$. This
property is called monogamy of entanglement. For an
infinite chain of qubits sharing uniform entanglement with their first
neighbors, Wootters \cite{Wootters} found an upper limit for concurrence
of $1/\sqrt{2}$ (this limit, however, has not been yet proven for
physical systems).  Under certain conditions, Wootters reports a
maximum concurrence of C$_{max}=0.434467$, a result which has also been
confirmed in the case of rings with $N$ qubits when $N \to \infty$
\cite{Connor}. D\"ur et al. \cite{Dur} have shown that in the case of
three qubits, average concurrence between pairs of qubits of the
$|W\rangle$ state are all equal to $2/3$, while Koashi et al.
\cite{Koashi} have reported a maximum average concurrence between all
pairs of qubits of $2/N$ when all qubits in the system except one are in
the state $|0\rangle$.

On the other hand, it has been conjectured by several authors that
entanglement can play an important role in quantum phase transitions
(QPT) \cite{Osterloh,Vidal,Gu,Yang,Wu,Larsson,Osborne}, which take place at absolute zero
temperature \cite{sachdev}. For example, Osborne et al. \cite{Osborne} have stated 
that in QPT, long-range correlations take place due to entanglement. 
In addition, Osterloh et al. \cite{Osterloh}  analyzed the behavior of entanglement near the critical point of
the spin $1/2$ model XY in a transverse magnetic field and  found that in the region close to a quantum
phase transition the derivatives of entanglement of formation (concurrence) 
obey a scaling law. Thus, there exists an intimate connection between entanglement,
scaling and universality. Further, quantum entanglement has
been used not only near quantum phase transitions to characterize
them, but also to obtain a better description of experimental measurements
of specific heat and magnetic susceptibility of dilute solution of
Ising dipoles \cite{Ghosh}.  Additionally, we have recently shown that, in the case of rings with 
off-diagonal disorder, an increase in disorder
strength results in {\it enhancement} of the nearest-neighbor (NN) concurrence with
respect to the perfectly ordered ring \cite{roman1}. This enhancement
of concurrence hints that quantum entanglement might be
a good indicator of anomalies in the wave functions and density of
states in systems where no quantum phase transitions take place.

In the present paper, we report maximization of the concurrence using
a computational optimization method. The calculations were done on the
basis of an electronic model Hamiltonian which contains parameters
depending on the lattice structure. Such parameters were optimized in
order to maximize the concurrence.  This means that, in general, the
systems considered will not be translationally invariant after
optimization, since the structure can be dramatically changed by the
genetic algorithm.  We present, specifically, calculations on systems
described by a tight-binding model in one and two dimensional
lattices. Our results can serve as a basis for the design of
structures which maximize nearest-neighbor entanglement.

This paper is organized in the following manner: in section II the
Hamiltonian is presented, as well as the most important formulas used
to quantify concurrence and the key steps in the optimization
procedure.  Results for one and two dimensional bipartite lattices, as
well as non-bipartite two dimensional lattices will be presented in
section III. Finally, a summary of our results can be found in section
IV.

\section{Theory} 
We consider electronic systems  described  
by a tight-binding Hamiltonian of the form 
\begin{equation}
\label{eq:Ham}
{\hat{H}}= \sum_{\langle ij \rangle}  t_{ij}   \hat c^{+}_i \hat c_j .
\end{equation}
For simplicity, we consider spinless electrons. In Eq.~(\ref{eq:Ham})
$\hat c^{+}_i$ ($\hat c_i$) is the usual creation (annihilation)
operator of a spinless electron at site $i$ and $t_{ij}$ is the hopping
integral between NN sites $i$ and $j$. 

It should be noted that the magnitude of the hopping elements $t_{ij}$ can be easily
related to the lattice structure. A large $t_{ij}$ indicates a small
interatomic distance, which causes a large overlap between the wave
functions localized on the sites $i$ and $j$. Analogously, a small
$t_{ij}$ represents a large interatomic distance. $t_{ij}=0$ means
that sites $i$ and $j$ are not nearest neighbors. 

To calculate the entanglement of formation between pairs of sites we
employed the concept of concurrence \cite{Wootters1} and proceeded
in a similar way as Zanardi and coworkers did for the case of
translationally invariant chains \cite{Zanardi}.  

For a system described by the Hamiltonian of Eq.~(\ref{eq:Ham}), the concurrence between sites 
$i$ and $j$ has the form\cite{Connor}
\begin{equation}
\label{eq:Conc}
{\rm C}_{ij}= 2 \max \{ 0, |z|-\sqrt{vy}\; \},
\end{equation}
where

\begin{align}
\label{eq:rhoaelement_v}
v &= 1- \langle \hat n_i  \rangle - \langle \hat n_j \rangle + 
\langle \hat n_i 
\hat n_j  \rangle \\
y &= \langle \hat n_i \hat n_j \rangle \\
z &= \langle \hat c^{+}_j \hat c_i \rangle .
\label{eq:rhoaelement_u}
\end{align}
Note that $i$ and $j$
do not need to be nearest neighbors. 
 
To maximize NN concurrence, we 
use genetic algorithms (GAs), a technique developed by J. H. 
Holland and his students at the University of Michigan in the 1960s and 
1970s. Holland's goal was not to design algorithms to solve specific
problems, but rather to formally study the phenomenon of adaptation as
it occurs in nature and to develop ways in which the mechanisms of
natural adaptation can be used for numerical optimization.

In this optimization technique, the characters or numbers representing
the solution to a problem are stored in a string called ``chromosome''.
Nature-inspired operators of crossover and mutation are applied to a 
population of chromosomes and those individuals that represent the 
best solution to the problem are given more probability of being chosen
for the next generation. By iterating this process, it is possible to obtain
a very good solution without having to explore the entire solution space.

For the particular problem treated in this paper, we propose first families of hopping integrals
$\{t_{ij} \}$, which are then fine-tuned by means of the
genetic algorithm.  Our goal is to obtain the hopping
elements from the Hamiltonian such that its ground state function
represents the maximum average NN concurrence. 

We have used genetic algorithms in a previous paper \cite{roman2}, 
where we have optimized the total concurrence
(C$_{total}$) of one dimensional chains and two dimensional lattices
described by a tight-binding Hamiltonian with open and periodic boundary 
conditions. C$_{total}$ was taken as an average over all concurrences 
between every site $i$ with the $N-1$ remaining sites. 

In this paper, however,  we focus only on
optimization of first-neighbor concurrence, which can be 
calculated by means of the following expression:
\begin{equation}
\label{eq:NNconc}
{\rm C}_{\rm NN}= \frac{1}{z N }\sum_{i=1}^{N} \sum_{j=1}^{z} {\rm C}_{ij}. 
\end{equation}
where $z$ is the number of nearest neighbors. C$_{\rm NN}$ is our
fitness function for the application of the genetic algorithm.

It is important to point out that a system with translational
invariance is described by all  the $t_{ij}$ elements from
Eq.~\ref{eq:Ham} having the same value, for example, $t_{ij}=-1$. This
configuration leads to concurrence values that are the same for each
pair of nearest neighbors, a case which has been previously studied by several
authors \cite{Connor,Zanardi,roman1}. 

The genetic pseudo-algorithm employed in this work has been given in detail 
elsewhere \cite{roman2}. Briefly, it consists in the following steps:
\begin{enumerate}
\item We consider each $t_{ij}$ of the Hamiltonian matrix as a gene and the 
array of these genes as a {\em chromosome}.
\item Allocate two arrays, ``generation0'' and  ``generation1''
 composed of chromosomes.
\item Allocate a chromosome ``best'' with fitness $0.0$.
\item For a given band filling, we repeat the following steps:
\begin{itemize} 
\item Initialize ``generation0'' with random values in the range $(-5,0)$.
\item For a given number of generations repeat: 
\begin{itemize} 
\item Decode each chromosome in ``generation0'' into a Hamiltonian matrix, 
diagonalize it and calculate the average concurrence between only 
nearest neighbors of the system using Eq.~\ref{eq:NNconc}. 
In other words, we calculate the fitness C$_{\rm NN}$ for each individual 
in ``generation0''. 
\item Compare ``best'' with the fittest individual in ``generation0'' and substitute ``best'' 
if the latter has larger fitness.
\item Choose chromosomes with a probability proportional to its fitness (selection operator) 
and copy it to ``generation1''.
\item Use crossover and mutation operators 
on chromosomes in ``generation1'' to create new chromosomes.
\item Make ``generation0'' equal to ``generation1''.
\end{itemize} 
\item Print ``best'' in an output file.
\end{itemize}
\end{enumerate}

\section{Results and Discussion} 
In this section, we present our results of NN concurrence optimization
using GAs. The effect of bipartite and
non-bipartite systems with periodic boundary conditions is 
addressed.  We recall that a lattice is considered bipartite if it can
be separated in two sublattices $A$ and $B$ such that all first
neighbors of the sites of sublattice $A$ are sites from sublattice $B$
and vice versa. Biparticity or non-biparticity has important
consequences in the physical properties of a lattice. For example, the
concurrence of a bipartite lattice is symmetric around half band
filling while non-biparticity is responsible for magnetic frustration
in spin systems, that is, the impossibility of minimizing energy for
each pair of spins in the lattice.

\subsection{Bipartite Systems}
\begin{figure}[thp]
\begin{center}
\resizebox{60mm}{!}{\includegraphics{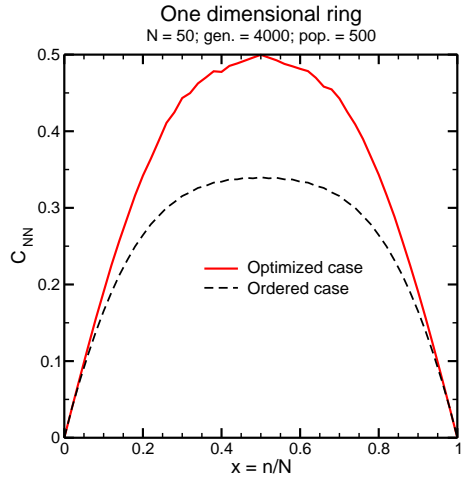}}
\caption{(Color online)
Nearest-neighbor concurrence C$_{\rm NN}$ of chains with periodic boundary conditions 
as a function of band filling $x=n/N$.}
\label{fig:1D}
\end{center}
\end{figure}

First, we focus on finite one dimensional chains of up to $N=50$ sites with periodic
boundary conditions.  
Results are shown in Fig.~\ref{fig:1D}. 
We considered 500 chromosomes (i.e. a collection of hopping integrals
from the hamiltonian matrix) in each population, with starting values randomly 
chosen in the range $(-5,0)$. The algorithm was left to evolve 
4000 generations in order to obtain the individuals with the best 
fitness value.
This procedure was undertaken for each band filling ($x=n/N$). From 
the figure it can be seen that there exists symmetry around half band
filling, indicating that it is possible to focus only on the $ 
0 \le x \le 0.5$ region 
when analyzing concurrence in bipartite lattices.

Optimized concurrence shows two distinctive features: (a) the growth
of concurrence as function of band filling is quasi-linear for $x \le 0.25$.
This result is due to the fact that in lower band fillings, the probability
of electronic collision is low, and the system is able to optimize concurrence
without considering Pauli repulsion. (b) For $x>0.25$ the growth of
C$_{\rm NN}$ as function of $x$ is quasi-parabolic, and electronic
 exchange due to the Pauli principle comes into play. Comparison between results 
 obtained using GAs and the ordered case 
--with all $t_{ij}$ taken equal-- shows that for low band filling ($x \le 0.1$),
 both concurrence curves display a similar behavior and the impact of optimization
starts to be noticeable for $x > 0.1$, with a maximum increase in concurrence
with respect to the ordered case of $47\%$ ($0.5$ vs $0.3392$) at half
band filling.

\begin{figure}[!ht]
\begin{center}
\includegraphics[width=0.9\textwidth]{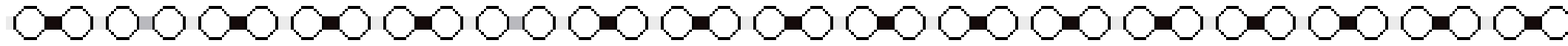}
\caption{(Color online) 
Graphical representation of a one-dimensional 
ring with $32$ sites in the lattice structure yielding  the maximal nearest-neighbor concurrence (see text). 
 Parameters: $5000$ generations;
 population=$600$. 
Half band filling. Three cases are distinguished by color: 
light grey ($|t_{ij}| < 1$), grey ($1 \le |t_{ij}| < 3$) and 
black ($3 \le |t_{ij}| < 5$).}
\label{fig:1D_x50}
\end{center}
\end{figure}

\begin{figure}[thp]
\begin{center}
\resizebox{80mm}{!}{\includegraphics{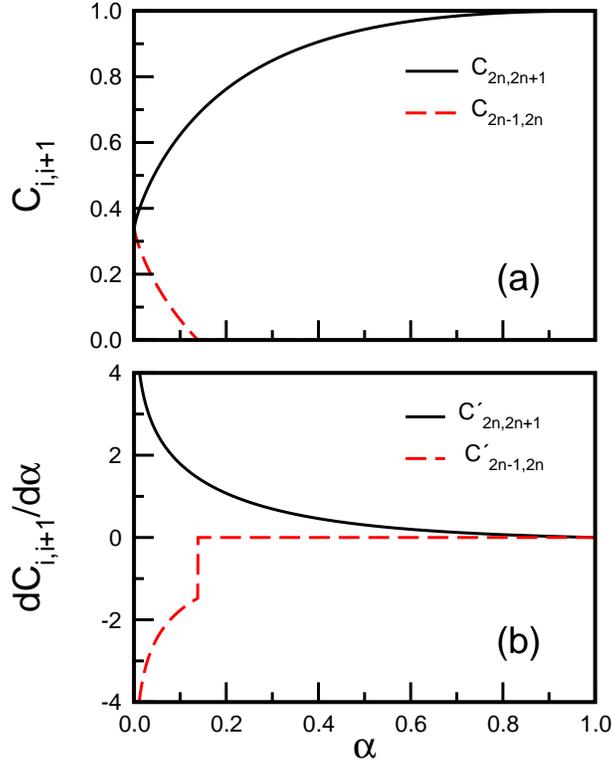}}
\caption{(Color online) (a)  Nearest-neighbor ground-state concurrences C$_{i,i+1}$  
of the dimerized chain (b) and their derivatives as a function of the dimerization parameter 
$\alpha$.}
\label{fig:Conc}
\end{center}
\end{figure}

In order to understand the behavior of optimized concurrence, we 
have analyzed those chromosomes with best fitness values for each
$x$. Results show that, as band filling $x$ is increased, the ring
starts to break into smaller chains (i.e., some hopping elements become $t_{ij} \simeq 0$). 
For example, while for $n=1$ all $t_{ij}$ are similar and close to $-4$,
in the next filling, $n=2$, two noncontiguous elements $t_{ij}$ differ from the rest and are now close to $0$,  
suggesting a separation into two subchains.
This division increases with $x$ until half band filling ($x=0.5$), a point at 
which the system alternates a short bond ($t_{ij} \simeq -5$) 
with a long one ($t_{ij} \simeq 0$) as can be seen depicted in Fig.~\ref{fig:1D_x50}. 
In other words, to increase the average concurrence,  the system begins to 
dimerize, i.e,  the chain  undergoes a structural phase transition due to a
 Peierls instability \cite{Peierls,Su,Fulde}.   This is a metal-insulator transition occurring in
one-dimensional metals, where the doubling of the unit cell leads to a
decrease in the kinetic energy of the system. Note that dimerization of the chain was 
obtained in a natural manner, since all  initial matrix elements were initialized by random values.
  In fact, the maximum concurrence at half band filling corresponds 
to a system which  has completely dimerized.  This shows a possible connection between the 
increase  in concurrence and the Peierls instability. In order to study this connection, 
we analyzed the concurrence and their derivatives for a dimerized chain. In this case, 
we consider  that the nearest-neighbor hopping integrals $t_{ij}$ of Eq.~\ref{eq:Ham} take the 
values $t_{2n,2n+1}=1+\alpha$ and $t_{2n-1,2n}=1-\alpha$, where $\alpha \in [0,1]$ is the 
dimerization parameter.  This  one-particle  dimerized Hamiltonian can be diagonalized by using the 
transformations proposed by Su, Schrieffer and Heeger \cite{Su}. 
The goal of analyzing the concurrence and its derivative is based on the fact that 
a discontinuity (singularity) in the  (derivative) ground-state concurrence has been 
associated to a first (second) order QPT \cite{Osterloh,Gu,Yang,Wu,Larsson}. 
However, it has been shown that the relation between QPT and non-analyticity
in the concurrence is not one-to-one  \cite{Yang}. A one-to-one  connection can be assumed though, 
when we consider QPTs characterized by nonanalytic behavior in the derivatives of the ground-sate 
energy and we exclude artificial and accidental occurrences of non-analyticities in the ground-state 
concurrence and its derivative \cite{Wu}.  In Fig.~\ref{fig:Conc}, we present the ground-state 
concurrence of a dimerized chain, C$_{2n,2n+1}$ and C$_{2n-1,2n}$,    
and its derivative, C$^{'}_{2n,2n+1}$ and C$^{'}_{2n-1,2n}$, where C$'$ means $d{\rm C}/d\alpha$, as a function of $\alpha$.  
Note that C$_{2n,2n+1}= 2.0*{\rm max}\{0, \gamma_{2n,2n+1}+
\gamma_{2n,2n+1}^2 - 0.25\}$ and C$_{2n-1,2n}= 2.0*{\rm max}\{0, \gamma_{2n-1,2n}+
\gamma_{2n-1,2n}^2 - 0.25\}$ \cite{roman1}, where $\gamma_{2n,2n+1}=\langle c^{\dagger}_{2n}c_{2n+1} \rangle$ 
and $\gamma_{2n-1,2n}=\langle c^{\dagger}_{2n-1}c_{2n} \rangle$  are  the one-particle density-matrix elements or bond 
orders between NN   and can be  calculated analytically \cite{Su}.  These bond orders  are continuous 
functions of $\alpha$, the first one ranging from $\gamma_{2n,2n+1}=0.318310$ ($\alpha=0$)
to 0.5 ($\alpha=1.0$) and the second one from  $\gamma_{2n-1,2n}=0.318310$  ($\alpha=0$)     
to 0.0.  ($\alpha=1.0$). Therefore,  the discontinuity obtained for  C$_{2n-1,2n}$ at $\alpha \approx  0.138$ is not 
related to  a critical point. Clearly, this discontinuity is  artificial  and  comes from the particular definition 
of the concurrence in Eq.~(\ref{eq:Conc}) \cite{Yang,Wu}.

In Fig. 3 one observes that C$^{'}_{2n,2n+1}$ and C$^{'}_{2n-1,2n}$ present singularities at the limit  
$\alpha \to \alpha_c=0$, where the Peierls instability occurs, which  should be related with 
a {\it second order QPT} \cite{Wu}.  In order to investigate if the singularity is related with a second order QPT, we write 
the ground state energy (and  their derivatives) as a function of  the NN density-matrix  elements (and their derivatives).  
The ground-state energy per site is given by  $E_{\rm gs}= -(1+\alpha)\gamma_{2n,2n+1}-(1-\alpha)\gamma_{2n-1,2n}$, 
its first derivative by $ dE_{\rm gs}/d\alpha= -(1+\alpha)d\gamma_{2n,2n+1}/d\alpha-(1-\alpha)d\gamma_{2n-1,2n}/
d\alpha-\gamma_{2n,2n+1}+ \gamma_{2n-1,2n} $
and  its second derivative by $d^2E_{\rm gs}/d^2\alpha= -(1+\alpha)d^2\gamma_{2n,2n+1}/d^2\alpha-(1-\alpha)d^2\gamma_{2n-1,2n}
/d^2\alpha -2 d\gamma_{2n,2n+1}/d\alpha + 2 d\gamma_{2n-1,2n}/ d\alpha$. 
These second derivatives presents a singularity at  $\alpha_c$,  which is a manifestation of a second order QPT. 

On the other hand, the maximal value of the C$_{\rm NN}$ is obtained when the chain is completely 
dimerized ($\alpha=1$), i.e. the chain has been transformed into $N/2$ singlet Bell states. 
This complete dimerization can be related to the ideas of monogamy between three systems \cite{Coffman}: 
the optimized system prefers to have maximum concurrence (C=1) between subsystems 
$A$ and $B$ while lowering that between $B$ and $C$ (C=0).

\begin{figure}[hbt]
\begin{center}
\resizebox{60mm}{!}{\includegraphics{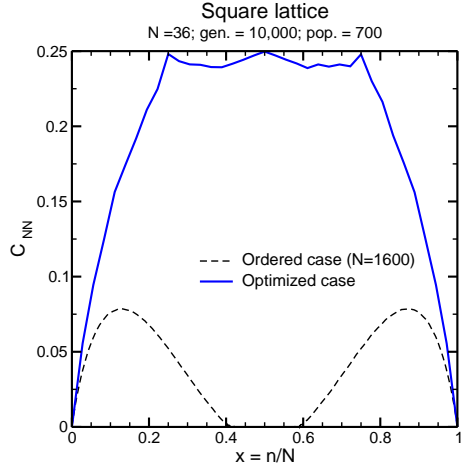}}
\caption{(Color online) 
Nearest-neighbor concurrence C$_{\rm NN}$ of a $6\times 6$ square lattice  
as a function of band filling $x$. The solid line refers to the optimized system. 
The dashed line shows the concurrence of a large perfectly periodic square lattice ($N=1600$).
}
\label{fig:C}
\end{center}
\end{figure}

The effect of dimensionality over C$_{\rm NN}$ is shown in
Fig.~\ref{fig:C}, where calculations for a square lattice $6
\times 6$ using a population of 700 individuals were performed. We considered 10,000
generations for each band filling.  From the figure, it is possible to
observe again two clearly remarkable traits for two ranges of $x$, as in
the one dimensional case: a low band filling range ($x \le 0.25$)
where the increase of C$_{\rm NN}$ as function of $x$ is quasi-linear due
to low collision probability between particles,
and a second range $x > 0.25$ where the Pauli exclusion principle starts to play an
important role.  C$_{\rm NN}$ as function of $x$ has a quasi-parabolic
behavior in the latter range, decreasing first in the range $0.25 < x \le
0.375$ and then increasing for $0.375 < x \le 0.5$.  It is also
noteworthy that both regions have the same maxima of C$_{\rm NN} \approx 0.25$, one at $x=0.25$ and
the other at $x=0.5$.  Comparison of C$_{\rm NN}$ between the optimized
case and an ordered, large lattice shows that the greatest difference occurs at half
band filling. Moreover, we observe that maxima  C$_{\rm NN}$ obtained 
for the square lattice is smaller than the maximum  C$_{\rm NN}$ for 
the 1D ring.

\begin{figure}
\centering
\subfigure[Band filling = 25\%]{\label{fig:C_x25}
\includegraphics[height=30mm]{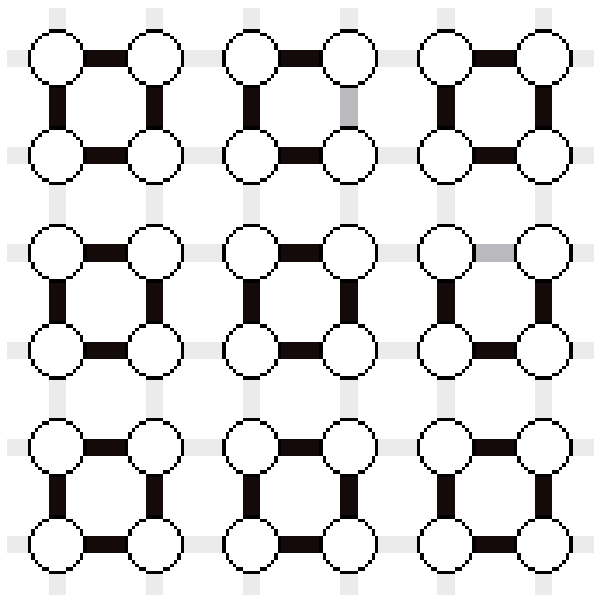}}
\qquad
\subfigure[Band filling = 50\%]{\label{fig:C_x50} 
\includegraphics[height=30mm]{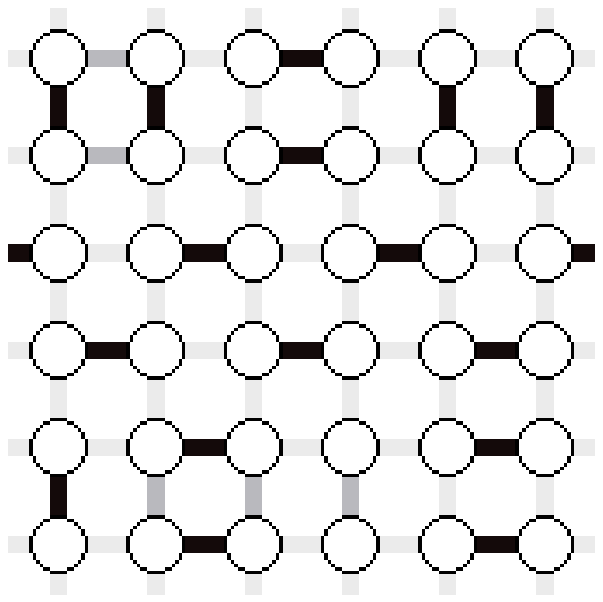}}
\caption{Graphical representation of the optimized structures obtained starting from a square lattice of 
36 $(6 \times 6)$ sites at (a) a quarter band filling and (b) half band filling. 
Parameters: $10000$ generations; population=$700$. Three cases are 
distinguished by color: light grey ($|t_{ij}| < 1$), grey ($1 \le |t_{ij}| < 3$) 
and black ($3 \le |t_{ij}| < 5$).}
\label{fig:C_dibujos}
\end{figure}

A study of the fittest chromosomes for those band fillings where
C$_{\rm NN}$ is a maximum shows the following behavior: for $x=0.25$, the
short bonds ($t_{ij} \simeq -5$) form nine squares which are
separated by long bonds ($t_{ij} \simeq 0$)
(Fig.~\ref{fig:C_x25}). In each square, a spinless
electron can be found. 
This  result is related to the
definition of generalized $| W_N \rangle$ states,
$|W_N\rangle = (1/ \sqrt{N})|N-1,1 \rangle$ where the state $|N-1,1
\rangle$ denote the totally symmetric state with $N-1$ zeros and one 1.
It has been shown that these $| W_N \rangle$ states are very robust
against particle losses and that the concurrence of two qubits can be
determined to be C$_{ij}=2/N$ \cite{Dur}.
In the case of $N=4$, we have that
$|W_4\rangle = (1/ \sqrt{4})\big[|0001 \rangle + |0010 \rangle + |0100
\rangle + |1000 \rangle \big]$ and  C$_{ij}$=0.5 for two NN sites of a
square. Using this C$_{ij}$ value for two NN sites of a square, it is
easy to show  that C$_{\rm NN}$=0.25 (see Eq.~\ref{eq:NNconc}).
On the other hand, for $x=0.5$ the system tries
once again to form dimerized states, separated with long bonds (Fig.~\ref{fig:C_x50}). 
The decrease of C$_{\rm NN}$ with respect to the one dimensional
ring is related to the difference in the number of 
neighbors between both systems. In the one dimensional systems, C$_{\rm NN}$ is
obtained by averaging between two nearest-neighbors whereas in the
square lattice this average requires four nearest-neighbors.

\subsection{Non Bipartite Systems}
In order to study the effect of non-biparticity  on the concurrence, we
have considered three non-bipartite lattices: the Kagom\'e
lattice, the maple leaf or Betts lattice and
the triangular lattice. Note that the triangular lattice is
less bipartite than the two other ones due to its larger number of
triangular bonds. This has important consequences in 
antiferromagnetic systems\cite{Aeppli}.

The basic common feature of the results presented so far is that
successful individuals in the linear chain and square lattices form
open and tightly closed binary subsystems. Thus, we exploited this fact
in order to speed up the procedure and increase the accuracy of the GA
calculations.  We initialized a hundred individuals of the population to
$0.0$ and a hundred to $-5.0$. The motivation for performing this step is  that this short and
long bonds will spread to other individuals through the crossover
operator, and that this kind of bonds will help increase its
concurrence.

\begin{figure}
\begin{center}
\resizebox{60mm}{!}{\includegraphics[height=30mm]{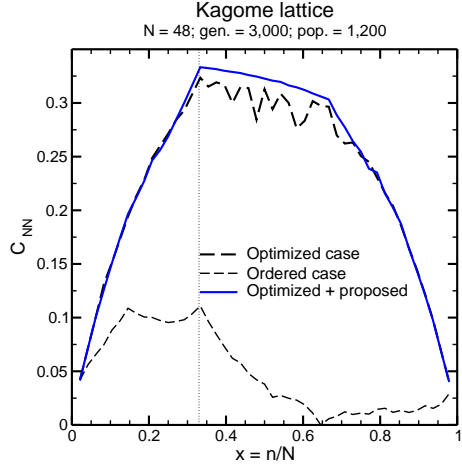}}
\caption{ (Color online)
Nearest-neighbor concurrence C$_{\rm NN}$ of small periodic 
Kagom\'e lattice as a function of band filling $x$. In addition, nearest neighbor concurrence 
C$_{\rm NN}$ for the Kagom\'e lattice were calculated using
the proposed optimal structure of  Fig.~\ref{fig:K_t}}
\label{fig:K}
\end{center}
\end{figure}

\begin{figure}
\centering
\subfigure[Best individual for Kagom\'e lattice]{\label{fig:K_x33}
\includegraphics[height=30mm]{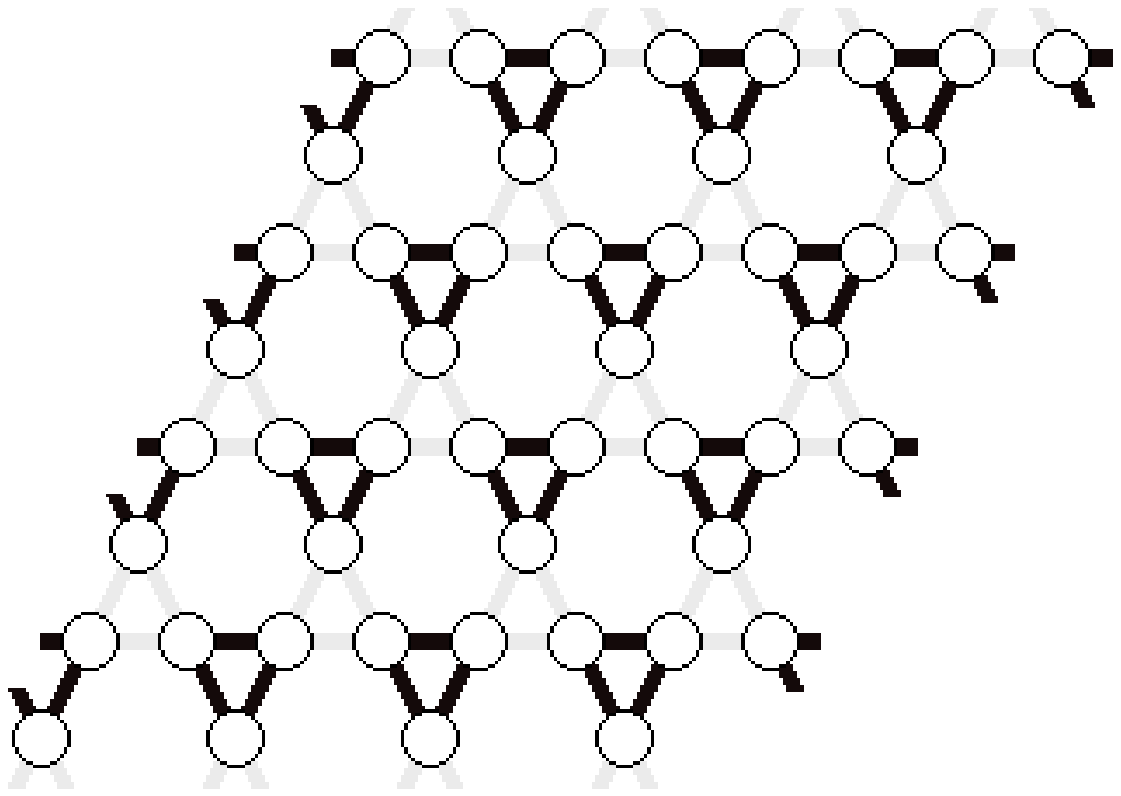}}
\qquad
\subfigure[Proposed optimal structure]{\label{fig:K_t}
\includegraphics[height=30mm]{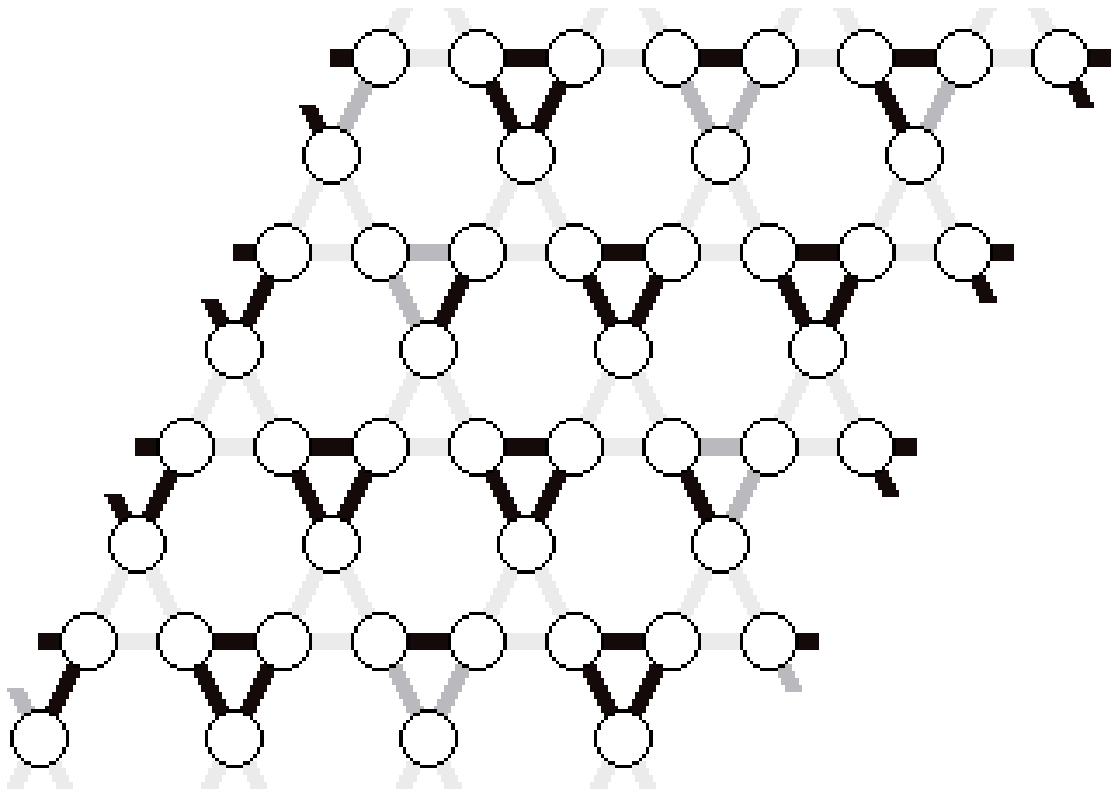}}
\caption{Graphical representation of a Kagom\'e lattice of $48$ sites. (a) 
represents the best individual using genetic algorithms at $x = 33\%$. 
(b) depicts a proposed optimal structure based on (a).}
\label{fig:K_dibujos}
\end{figure}

We now apply our calculations to a periodic Kagom\'e lattice of 48 sites with 
 periodic boundary conditions. Results are shown in Fig.~\ref{fig:K}.
 We considered 1200 individuals
in the population, which was left to evolve over 3000 generations. 
Quasi-linear behavior of C$_{\rm NN}$ with $x$ can be seen in regions with low
electronic or hole density, namely $x \le 0.2$ and $x \ge 0.8$. This
behavior corresponds to low probability of collision, as has already
been discussed, and is independent from the kind of lattice or whether
it presents triangular bonds or not. The first effect of non-biparticity
is reflected in the lack of symmetry of C$_{\rm NN}$ around half band filling.
Note that Fig.~\ref{fig:K} presents only one distinguishable maximum
located --contrary to the case bipartite lattices--, out of half band filling,
 at $x \simeq 1/3$.
Comparison with respect to the ordered case shows the relevance of the optimization
technique,  particularly for $0.6 \le x \le 0.8$, 
where the initially ordered structure  displays a very low concurrence. Moreover, there is a
coincidence in the localization of the maximum for both the optimized
and ordered lattices.
The best individual corresponding to the maximum of the numerical optimization
approach can be seen in Fig. \ref{fig:K_x33}, where one can observe that a geometrical
configuration favoring triangular loops has been formed. It is clear 
again that this trend to form triangular loops is related to $W$ states,
 $|W \rangle = (1/\sqrt{3}) \big[ |100\rangle + |010\rangle +   
|001\rangle  \big]$,  where C$_{ij}=2/3$ for two NN sites forming the
triangle.  To further extend the optimized
results, we now introduce a design as seen in Fig. \ref{fig:K_t}.
A new concurrence curve for the Kagom\'e lattice is presented in Fig. \ref{fig:K}. In 
this case, besides the hundred individuals initialized to $0.0$ and $-5.0$ each, all
other individuals were initialized with the proposed design of Fig. \ref{fig:K_t}.
As can be seen, the concurrence is not only increased at the maximum, but the 
proposal improved the overall behavior of the curve as well.

\begin{figure}
\centering
\subfigure[Best individual for Betts lattice]{\label{fig:B_x33}
\includegraphics[height=40mm]{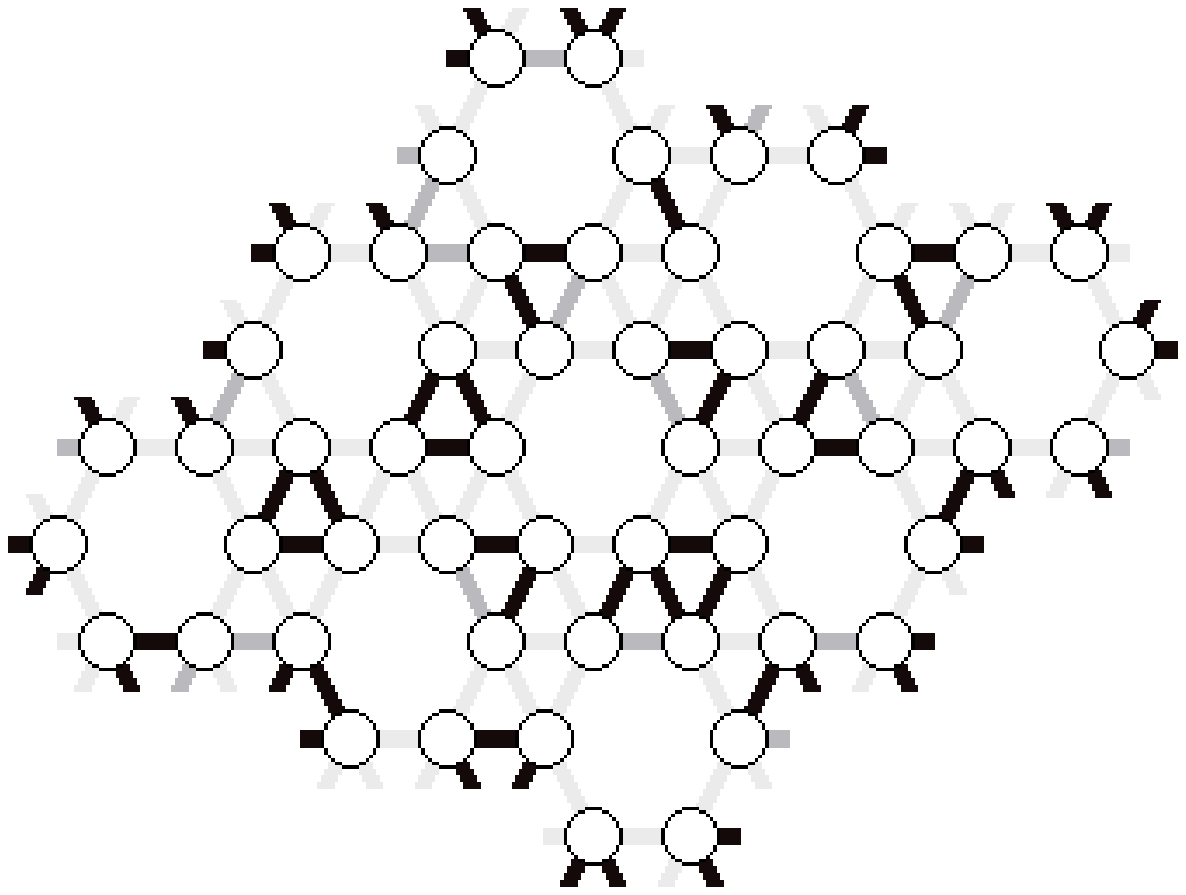}}
\qquad
\subfigure[Proposed optimal structure]{\label{fig:B_t}
\includegraphics[height=40mm]{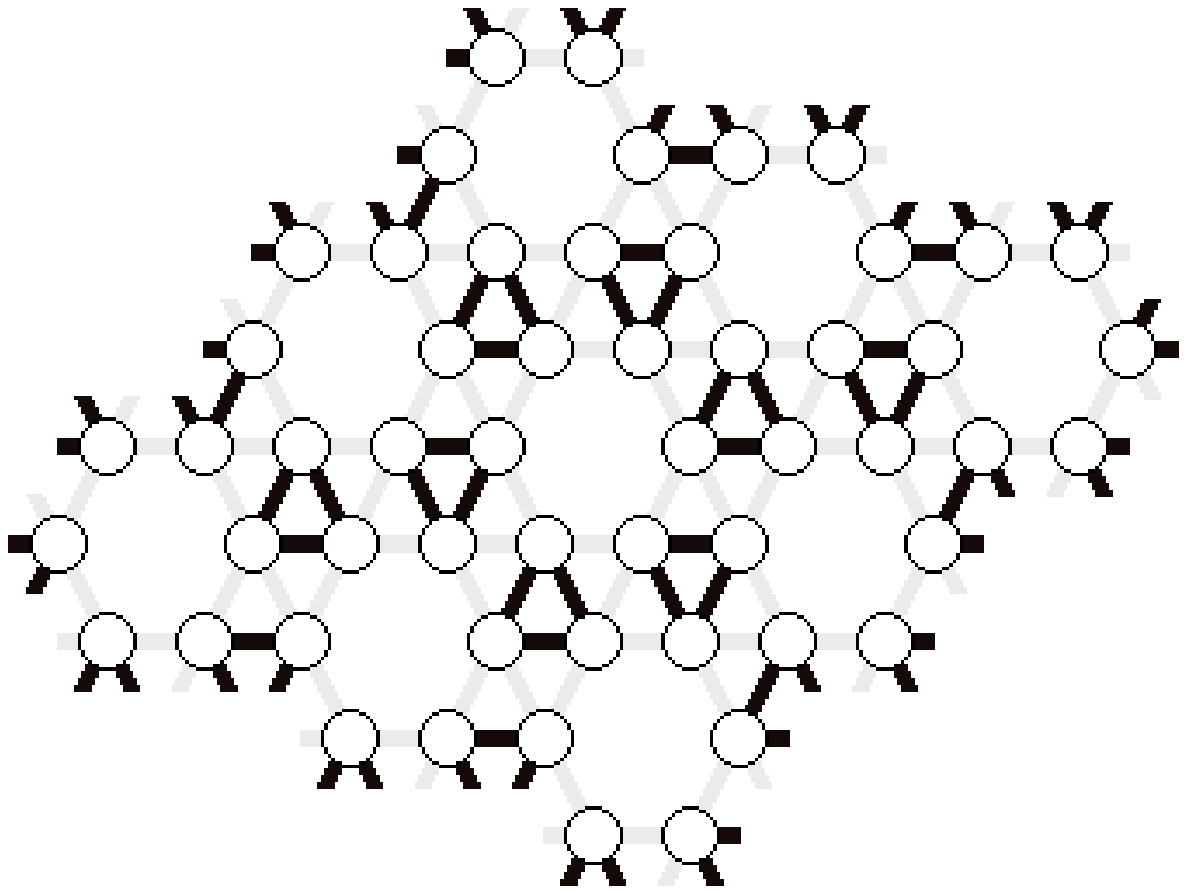}}
\caption{Graphical representation of a Betts lattice of $48$ sites. (a) 
represents the best individual using genetic algorithms at $x = 33\%$. 
(b) depicts a proposed optimal structure based on (a).}
\label{fig:B_dibujos}
\end{figure}

\begin{figure}
\begin{center}
\resizebox{60mm}{!}{\includegraphics[height=30mm]{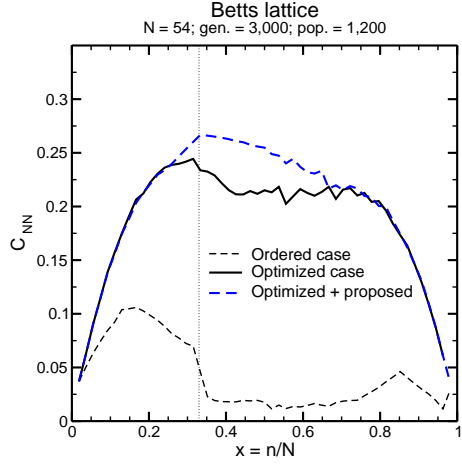}}
\caption{ (Color online) 
Nearest-neighbor concurrence C$_{\rm NN}$ for the Betts 
lattice as a function of $x$. The figure shows C$_{\rm NN}$ for the ordered case, the optimized 
case using genetic algorithms and, finally, the case where the generations were initialized 
at each band filling using
the proposed optimal structure of Fig.~\ref{fig:B_t}.}
\label{fig:B}
\end{center}
\end{figure}

Considering the hint given by the analysis of  the Kagom\'e lattice, we ran the optimization algorithm for
the maple-leaf or Betts lattice in order to find an optimum structure design. Fig. \ref{fig:B_dibujos} 
shows the best individual of an initial run of the algorithm corresponding
to the maximum at $x=0.33$ [Fig. \ref{fig:B_x33}], as well as a proposal based on this 
individual [Fig. \ref{fig:B_t}]. 
Fig. \ref{fig:B} shows the results for the ordered case, as well as both runs 
of the algorithm. As expected, the ordered case performed worse than any of the optimized
cases. Once again, the run based on proposal shown in Fig. \ref{fig:B_t} performed 
better than the random-based case.

\begin{figure}
\begin{center}
\resizebox{60mm}{!}{\includegraphics[height=30mm]{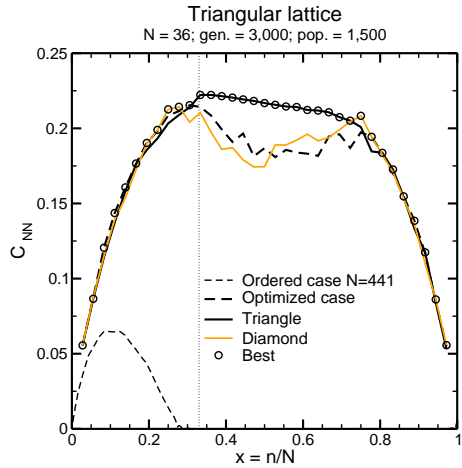}}
\caption{(Color online) 
Nearest-neighbor concurrence C$_{\rm NN}$ for the triangular lattice as 
a function of $x$.  The nearest-neighbor concurrence C$_{\rm NN}$
was calculated  for the ordered case, the optimized case using genetic algorithms, 
the proposed optimal structures and the case combining results from proposals.}
\label{fig:T_mejor}
\end{center}
\end{figure}

\begin{figure}
\centering
\subfigure[Proposed optimal structure ``Triangle'']{\label{fig:T_t2}
\includegraphics[height=30mm]{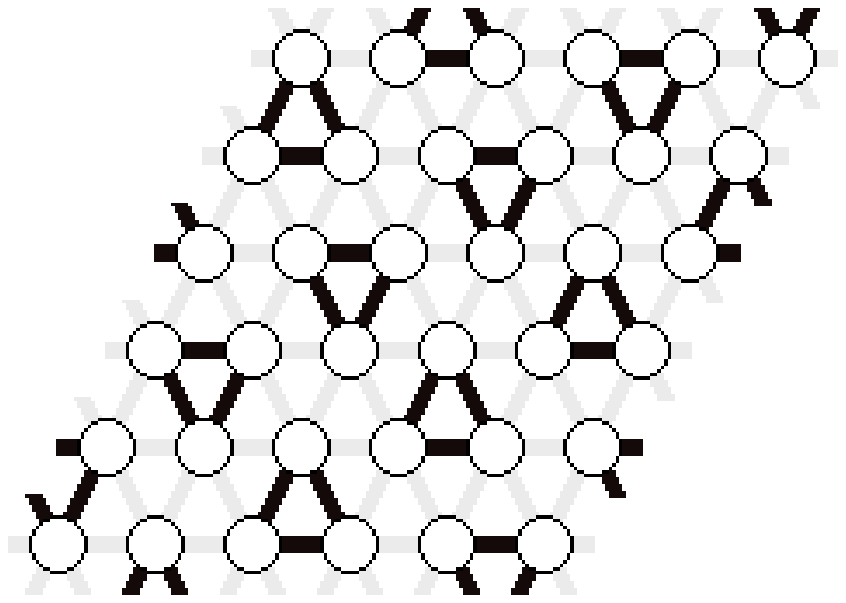}}
\quad
\subfigure[Proposed optimal structure ``Diamond'']{\label{fig:T_d}
\includegraphics[height=30mm]{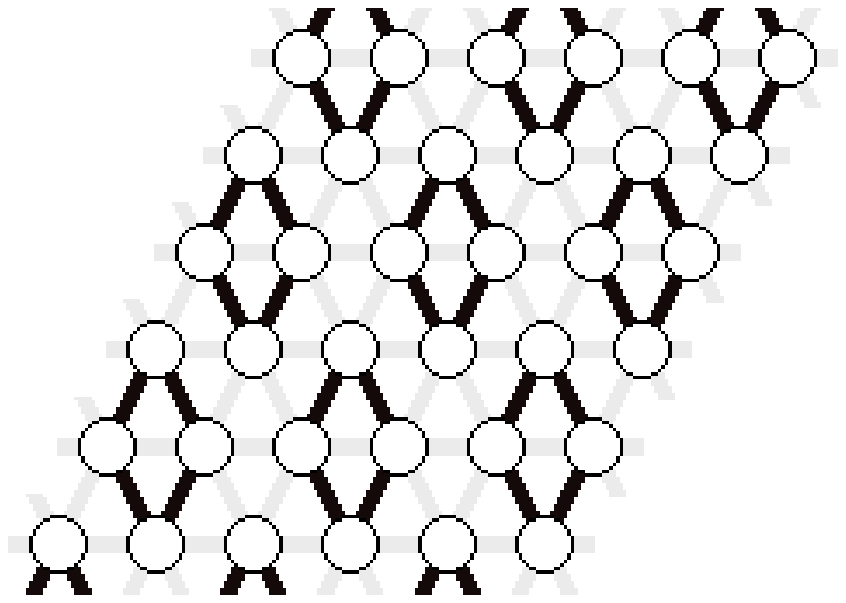}}
\caption{Proposed optimal structures based on the best individual obtained 
using genetic algorithms. The “triangle” structure is based on 
the best individual at $x = 33\%$ whereas the diamond is inpired on the best 
individual at $x=0.69$.}
\label{fig:T_propuestas}
\end{figure}

Finally, in the case of the triangular lattice, the typical behavior found in the other 
non-bipartite lattices can be readily observed: a single maximum located near $x=1/3$ and a much better
performance with respect with the ordered case (see Fig.~\ref{fig:T_mejor}).
As in the other cases, we then proceeded to develop new proposed configurations by looking
at successful individuals and being inspired by $W$ states. Two new designs were chosen, 
seen in Figures  \ref{fig:T_t2} and \ref{fig:T_d}.

The best curve for C$_{\rm NN}$ can be seen in Fig. \ref{fig:T_mejor}, and
is  basically a combination of the different curves obtained when initializing with 
each proposal. Although there is a fair increase in the overall maximum
 (C$_{\rm NN}=0.2222$), it is still lower than that of the 
Kagom\'e lattice (C$_{\rm NN}=0.3333$) and maple leaf lattice (C$_{\rm NN}=0.2666$). 
This is because in the triangular lattice there are more bonds, and consequently each site 
must share its entanglement with more neighbors.

\section{Summary and conclusions}
In this work, we have used  GAs to maximize the nearest-neighbor
average concurrence of systems by tuning the nearest-neighbor hopping
integrals of a tight binding Hamiltonian. The optimization of
entanglement has been performed for one-  and two dimensional
bipartite systems as well as for two dimensional non bipartite systems. The
results show that the concurrence of the optimized systems is very
large in comparison with the ordered structures. This
increase in the concurrence is understood and interpreted by analyzing the optimized
nearest-neighbor hopping integrals.  In general, we found certain
tendencies of periodical systems to break into smaller subsystems. 
This is achieved in a natural manner by the system by making the hopping
integrals evolve in such a way that the absolute value of some 
integrals is high in some cases ($|t_{ij}| \simeq 5$) and very small in 
others ($t_{ij} \simeq 0$).

This results are related to the fact that quantum entanglement, in contrast
with classical correlations, cannot be freely shared between many objects.
This quantum correlations property ---monogamy---, is clearly noticeable
in the case of the periodic ring at $x=0.5$, where the $n$-th site is completely
entangled with the site $n+1$ (C$_{\rm n, n+1} \simeq 1$) while on 
the other hand it is completely unentangled with the site $n-1$ 
(C$_{\rm n, n-1} \simeq 0$). Moreover, results at $x=0.5$ shows that, 
in order to maximize concurrence, the system undergoes a structural
 transition, the Peierls distortion.

Finally, it is worth mentioning that the relationship between
the 1D tight binding Hamiltonian [Eq.~(\ref{eq:Ham})]  and 
the $XX$ chain of 1/2 spin given by the following expression 
\begin{equation}
\hat H_{XX}=\sum_j^{N_a} \frac{J_j}{2} 
( S^{+}_{j+1} 
S^{-}_j +  S^{-}_j 
S^{+}_{j+1} ),
\end{equation}
where ${J_j}$ is the coupling constant. 
This relationship can be shown by using the Jordan-Wigner
 transformation \cite{sachdev} and therefore, it can be 
inferred that our results for the 1D tight binding systems can be used 
to describe the maximization of the C$_{\rm NN}$ concurrence for systems 
modeled by the $XX$ chain of 1/2 spin.

\begin{acknowledgments} 
We thank CONACYT for providing support for one of the authors
(J.C.N.M).  
The computational resources that we used for this work were provided by
the National Center of Supercomputing CNS-IPICyT, Mexico.
\end{acknowledgments}

\end{document}